\documentclass[conference]{IEEEtran}
\IEEEoverridecommandlockouts
\usepackage{amsmath,amssymb,amsfonts}
\usepackage{algorithmic}
\usepackage{graphicx}
\usepackage{csquotes}
\usepackage{textcomp}
\usepackage{ulem}
\usepackage{xcolor}
\def\BibTeX{{\rm B\kern-.05em{\sc i\kern-.025em b}\kern-.08em
    T\kern-.1667em\lower.7ex\hbox{E}\kern-.125emX}}
\usepackage{hyperref}

\begin{document}

\title{Methods To Ensure Privacy Regarding Medical Data\\
\vspace*{0.6\baselineskip}
{\huge Including an examination of the differential privacy algorithm RAPPOR and its implementation in \enquote{CrypTool 2}}
}

\author{\IEEEauthorblockN{Christina Wölk}
\IEEEauthorblockA{\textit{University of Siegen} \\
\textit{Student (Bachelor's Degree in Computer Science)}\\
christina.woelk@student.uni-siegen.de}
}

\maketitle

\begin{abstract}
This document examines several applicable methods to ensure privacy of data gathered in the health care sector. To ensure a common understanding of the topic, the introduction explains the need for anonymization methods based on an example. Next, reasons for data collection are introduced in connection to the purpose to protect mentioned data, as well as currently applicable privacy laws to enforce this privacy. 

The question \textit{What kind of privacy we are talking about and what conditions have to be fulfilled} is dealt within the subsequent chapter \enquote{\nameref{sec:Differential_Privacy}}.
Thus being established, common anonymization methods are explained and reviewed for their use in the healthcare sector.

The RAPPOR algorithm and its differential privacy is dealt with in more detail before coming to a conclusion.

\end{abstract}

\vspace*{0.6\baselineskip}
\section{Introduction}
Privacy is valued by the majority of German citizens. With the increasing amount of possibilities to use technology and collect data, safety measures have to be increased as well. So-called "anonymity" (in this case meaning censoring one's name in a dataset) alone is not sufficient anymore to ensure one's privacy. We often can be identified by a very limited amount of data (e.g., date of birth, hometown and gender). Since personal data is often collected in several matters and from several sources, records (e.g., medical records) can be allocated to a specific person and exposed. An example from the 1990s:\\

The \enquote{Massachusetts Group Insurance Commission} released anonymized data on state employees revealing hospital visits to establish a bigger database for researchers. During the anonymization, name, address and social security number were removed. Latanya Sweeney (at that time a graduate student in computer science) was able to identify the then Governor of Massachusetts just by buying the voter rolls (for approximately 20 US-Dollars\cite{governor})  from the city of Cambridge (where she knew the governor resided) which included name, address, ZIP code, birth date and sex of every voter. Now knowing his ZIP code, she was able to identify his medical records; and therefore, knowing every diagnosis and every prescription. \cite{governor}\\

This example depicts the limitations of anonymization. In this case, the student left it at sending the medical records to the governor's office \cite{governor}. However, so-called "linkage attacks" (where sensitive information can be allocated to the person it belongs to) still are a risk that should not be ignored. Therefore, this paper examines methods to ensure privacy regarding medical data, especially focusing on the RAPPOR algorithm provided by Google, as described in the abstract.

\vspace*{0.6\baselineskip}
\section{Data Protection in the Healthcare Sector}
Information collected for research purposes is not new. However, when dealing with medical researches, certain problems occur to a greater length then elsewhere. The most frequent one is the problem of the size of the study. When researching how to improve health, the need for probands suffering from certain medical conditions is almost inevitable. Even something small as a new medicine that alleviates headaches can only be tested on persons who have a headache for obvious reasons. For rare diseases, this can become quite the big problem since the results would not be significant. So the group of potential probands is often very limited and can not be enlarged (also for obvious, ethical reasons). This complicates the anonymization because the larger the group, the harder it is to identify the individual.

\vspace*{0.6\baselineskip}
\subsection{What kind of data is gathered? (And for what reasons?)} 
Data is not always the same. Especially in the health care sector, the patients data is often very sensitive. Let us review the data types we often deal with in the following sections.\\

\subsubsection{Data concerning age and sex}
Effects and side effects of medical treatments can differ depending on the sex of the patient. This correlates with height and weight influencing the appropriate dose rate, as well as the difference of hormones of the patient interacting with specific medications. \cite{gendermedicine}
Therefore this information is needed to ensure a proper medication. But not only is this relevant for treatment, it is also essential for the prevention of statistical probable illnesses or diseases. A good example is the invitation of the general practitioner to get a checkup for breast cancer when a female reaches the age of 30 or the checkup for prostate cancer when a male reaches the age of 45 \cite{vorsorge}. Even if the causes for the increased occurrence happen to be unknown, taking precautions is recommended because pure statistics do, in fact, save lives.\\

All data regarding the treatment, as well as the diagnosis, are protected by medical confidentiality and data protection ordinance which the patient has to agree upon.\\

\subsubsection{Socioeconomic factors}
Socioeconomic factors such as income, spoken languages, place of residence, job and family status are rarely relevant for ordinary treatment. Nevertheless they can be of utmost importance in terms of research, especially long term studies. The development and emergence of many diseases aren't dependent on one, but a large amount of factors. For instance, it is known that there are many factors that increase the risk of getting cancer at a younger age (although, there still is a lot of research to be done). Therefore, this kind of data mining is necessary to get closer to gain insight on these matters.\\

\subsubsection{Data regarding lifestyle}
Balanced diet, doing sport on a regular basis, etc., influences health in a positive way; to the contrary, a lack of exercise and eating a lot of unhealthy food increases the risk of diseases like type 2 diabetes \cite{diabetes}. This data is also important for long term studies, as well as almost everything else, as it affects our everyday life. It is hard to measure and is almost solely observed by the patient rather than the doctor. This leads to inaccuracy which is why a lot of data has to be gathered before it can be deemed useful.\\

\vspace*{0.6\baselineskip}
\subsection{What are the consequences of insufficient protection of medical data?}

\subsubsection{Patient's point of view}
As can be seen in the paragraphs above, medical data consists of a variety of sensitive data concerning our private life. Many of these factors (e.g., income) indicate or influence a certain social status. Revealing sensitive data can lead to social stigmatization, and therefore mental stress. A doctor or researcher will not judge you for having mental illnesses or consuming drugs, but your social environment or your employer might do so. It does not have to be something with a lot of prejudices attached to it. The concern of you, as an employee, who is on sick leave more often due to being a migraine patient might be enough for you to not get the job.\\

Other mentionable parties which are potentially interested in your medical data are insurance companies. They want to know the risk of having to pay for you as their customer or having a reason to make the insurance more expensive for you. Additionally, it has to be considered that information regarding one's medical data can also affect other people, as some diseases are genetic.\\

\subsubsection{Company's point of view - The General Data Protection Regulation} 
Since 2018 the General Data Protection Regulation based on a decree of the European Union is in force in Germany \cite{dataRegulation}. Its main goal is to ensure informational self-determination as well as other fundamental freedoms.
In Germany, the person to whom the data is attributive is the owner of the data. Therefore, it is prohibited to assimilate personal data unless stated otherwise. Data protection aims for data integrity, data confidentiality and (for this topic especially important) data resilience which means the resilience towards, e.g., hackers. \cite{dataRegulation}\\

Violations of the General Data Protection Regulation can lead to fines up to 20 million Euro or 4\% of the worldwide sales of the company (depending on which amount is higher) \cite{dataRegulation}.

\vspace*{0.6\baselineskip}
\section{Differential Privacy}
\label{sec:Differential_Privacy}

Differential privacy pursues the goal to obtain as accurate responses as possible (e.g., from surveys or user behaviour) while making it as difficult as possible to identify a person by his or her given answers. The parameter $\epsilon$ is used to "measure" the extent of the given privacy. A small $\epsilon$ represents a high privacy guarantee, as a consequence of the definition of differential privacy:

"A randomized function $\kappa$ gives 	$\epsilon$-differential privacy if for all
data sets D1 and D2 differing on at most one element, and all $S \subseteq Range(\kappa)$: 

\begin{equation}  
{\Pr[\kappa (D_{1})\in S]\leq e^{\varepsilon }\times \Pr[\kappa (D_{2})\in S]}.
\end{equation}"\cite{privacy}. Range($\kappa$) is the set of every possible outcome of function $\kappa$ which could, for example, be the set of all whole numbers. The left side of the inequation describes the probability (Pr) that the full database (D1), randomized by the function $\kappa$, is included in the subset ($S$). The right side does the same except one entry has been removed from the database and the term is multiplied ($\times$) with $e^\epsilon$. Note that for $\epsilon=0$, the term is multiplied with one, giving the highest possible privacy. This means that the privacy for each user is about the same and it does not matter whether a person is included in the database or not.
When using differential privacy methods, the real responses aren't necessarily sent to the server. Instead, with a certain probability, the given answer will be a random one. This protects the users data, even if the users response is intercepted multiple times, because the real response is harder to reconstruct. Differential privacy is not focused on the method, but on the result: how well is the privacy of the user protected?

\vspace*{0.6\baselineskip}
\section{Overview of privacy preserving methods}
Data mining in the health care sector can improve, e.g., detection of diseases, but requires to ensure the patients privacy.\\

Anonymization in general aims to cover up personal or delicate information concerning the proprietor. This information can be classified into four different types of information\cite{currentMethods}:
    \begin{itemize}
        \item \textbf{Explicit Identifiers} - for example, name or a unique id can identify a person without further information.
        \item \textbf{Quasi-identifiers} - like age, sex, or ZIP code can identify a person via combination with different attributes as explained in the introduction.
        \item \textbf{Sensitive Attributes} - confidential and/or personal information about the patient as, for example, a diagnosis or prescription.
        \item \textbf{Non-Sensitive Attributes} - information which does not require anonymization. Mentioned for the sake of completeness but irrelevant for our topic.
    \end{itemize}
    
    \begin{table}[h!tbp]
    \caption{Data set before Anonymization}
    \begin{center}
    \begin{tabular}{|c|c|c|c|c|}
    \hline
    \multicolumn{5}{|c|}{\textbf{Extract of Medical Record (Fictional)}} \\
    \cline{1-5}
    \textbf{\textit{Name}}& \textbf{\textit{Age}}& \textbf{\textit{Gender}}& \textbf{\textit{ZIP Code}} &\textbf{\textit{Diagnosis}} \\
    \hline
      Jane Doe& 44 & Female & 12345 & Cancer  \\
    \hline
        John Smith & 22 & Male & 12333 & Migraine \\
    \hline
        William Wonker & 39 & Male & 12344 & Incontinence\\
    \hline
     Harrison Seat & 35 & Male & 12355 & Incontinence\\
    \hline
     Bettina Wonker & 42 & Female & 12344 & No illness\\
    \hline
     Thomas Müller & 22 & Male & 12222 & Diabetes\\
    \hline
     Sharon Carter & 47 & Female & 12544 & Cancer\\
    \hline
     Maria Granger & 27 & Female & 12345 & Cancer\\
    \hline
     Christian Cloud & 26 & Male & 12333 & No illness\\
    \hline
     Kim Schmidt & 21 & Female & 12222 & Diabetes\\
    \hline
    \end{tabular}
    \label{tab1}
    \end{center}
\end{table}
 
 In the example above, the name is an Explicit Identifier whereas Age and Gender are Quasi-Identifiers. Needless to say, the diagnosis contains the sensitive information and the data set contains a lot more entries than the ten exemplary above.
 The acquired data can be anonymized in different ways which are described in the following.

\vspace*{0.6\baselineskip}
\subsection{Basic anonymization methods}
    
    \subsubsection{Generalization and suppression}
    One way to anonymize data is to remove (or "suppress") Explicit Identifiers and generalize or suppress Quasi-identifiers which makes it more difficult to identify a person in the data set. Values that can not be suppressed completely are, instead, replaced with the category they can be put into \cite{gensur}. For example, instead of an age of 22, only the span from 20 to 30 is given.

 \begin{table}[h!tbp]
    \caption{Data set after generalization and suppression}
    \begin{center}
    \begin{tabular}{|c|c|c|c|c|}
    \hline
    \multicolumn{5}{|c|}{\textbf{Extract of Medical Record (fictional)}} \\
    \cline{1-5}
    \textbf{\textit{Name}}& \textbf{\textit{Age}}& \textbf{\textit{Gender}}& \textbf{\textit{ZIP Code}} &\textbf{\textit{Diagnosis}} \\
    \hline
      * & $40 \leq age \leq 49$ & Female & 12* & Cancer  \\
    \hline
        * & $20 \leq age \leq 29$ & Male & 12* & Migraine \\
    \hline
        * & $30 \leq age \leq 39$ & Male & 12* & Incontinence\\
    \hline
     * & $30 \leq age \leq 39$ & Male & 12* & Incontinence\\
    \hline
     * & $40 \leq age \leq 49$ & Female & 12* & No illness\\
    \hline
     * & $20 \leq age \leq 29$ & Male & 12* & Diabetes\\
    \hline
     * & $40 \leq age \leq 49$ & Female & 12* & Cancer\\
    \hline
     * & $20 \leq age \leq 29$ & Female & 12* & Cancer\\
    \hline
    * & $20 \leq age \leq 29$ & Male & 12* & No illness\\
    \hline
    * & $20 \leq age \leq 29$ & Female & 12* & Diabetes\\
    \hline
    \end{tabular}
    \label{tab22}
    \end{center}
\end{table}

The anonymized data can be structured into several equivalence classes. Within an equivalence class (for example, male patients between 20 and 30 years old living with ZIP codes starting with 12) every patient is indistinguishable. To indicate the privacy, one examines the k-anonymity in which k is the minimum amount of people in an equivalence class \cite{kAnonymity}. Therefore, a large k indicates a high difficulty when trying to identify a person in the data set. However, as can be seen in Table \ref{tab22}, the given privacy comes at the cost of information loss.

\begin{table}[h!tbp]
    \caption{Data set sorted into equivalence classes}
    \begin{center}
    \begin{tabular}{|c|c|c|c|c|}
    \hline
    \multicolumn{5}{|c|}{\textbf{Extract of Medical Record (fictional)}} \\
    \cline{1-5}
    \textbf{\textit{Name}}& \textbf{\textit{Age}}& \textbf{\textit{Gender}}& \textbf{\textit{ZIP Code}} &\textbf{\textit{Diagnosis}} \\
    \hline
    \multicolumn{5}{|c|}{Equivalence class one}\\
    \hline
      * & $40 \leq age \leq 49$ & Female & 12* & Cancer  \\
    \hline
     * & $40 \leq age \leq 49$ & Female & 12* & Cancer\\
    \hline
     * & $40 \leq age \leq 49$ & Female & 12* & No illness\\
    \hline
    \multicolumn{5}{|c|}{Equivalence class two}\\
    \hline
     * & $20 \leq age \leq 29$ & Female & 12* & Cancer\\
    \hline
    * & $20 \leq age \leq 29$ & Female & 12* & Diabetes\\
    \hline
    \multicolumn{5}{|c|}{Equivalence class three}\\
    \hline
        * & $20 \leq age \leq 29$ & Male & 12* & Migraine \\
    \hline
    * & $20 \leq age \leq 29$ & Male & 12* & No illness\\
    \hline
     * & $20 \leq age \leq 29$ & Male & 12* & Diabetes\\
    \hline
    \multicolumn{5}{|c|}{Equivalence class four}\\
    \hline
        * & $30 \leq age \leq 39$ & Male & 12* & Incontinence\\
    \hline
     * & $30 \leq age \leq 39$ & Male & 12* & Incontinence\\
    \hline
    
    \end{tabular}
    \label{tab3}
    \end{center}
\end{table}

The k-anonymity in the table above equals two because in every equivalence class there are at least two individuals which are indistinguishable from another.
 This k-anonymity does not guarantee that sensible information can not be revealed (even a higher k-anonymity does not do so). Under the assumption that one knows a person must be in the data set (since one knows he/she participated in the study), one can easily determine the equivalence class the person must be in, via traits, which are either known or obvious (gender, age, etc.). If every entry in the concerning equivalence class contains the same sensible data (like it is in the case in equivalence class four of our example) one still gains sensible information about the patient. Especially in studies with cases in which the disease seems to correlate to age, this issue is not to be ignored.\\

K-Anonymity can be extended by another indicator of privacy: \textbf{l-diversity}. L-diversity states how diverse an equivalence class (a class containing only entries which have are the same in the matter of defined attributes) is. Formally, l-diversity is defined as followed:\\

\textit{"Let a q*-block be a set of tuples such that its non-sensitive values generalize to q*. A q*-block is l-diverse if it contains l "well represented" values for the sensitive attribute S. A table is l-diverse, if every q*-block in it is l-diverse."} \cite{ldivers}\\

This addresses exactly the issue explained above. When maintaining l-diversity, it is guaranteed that the sensitive information is not revealed by simply figuring out the persons equivalence class because now each class has to contain $l$ different values for the sensitive information. However, l-diversity also has its limits. The values can be similar to each other even when not being the same, for example, "alimentary gastritis" and "anacide gastritis" or "blood cancer" and "skin cancer". A class containing only cancer patients, even if everyone has a different type of cancer, can still lead to the patient being identified as a cancer patient.\\

Using (only) generalization and suppression for medical data can be quite difficult because the accuracy of the data (e.g., the age) can be essential to obtain usable results for many medical researches. Especially with the often limited amounts of probants, every loss of information can be critical. Therefore, generalization is not always an option. However, if the accuracy of the given data is not essential to the usability of the results, anonymization is an uncomplicated and easily implemented strategy to ensure privacy. This could be the case when only general amounts are observed. For example, if a researcher is looking for the amount of alcohol consumed by young people compared to older ones, it is not critical to know the exact age.\\

\subsubsection{Noise addition}
Noise addition is another method which alters the given data set. The noise consists of random values that replace the original values. In our context it makes sense to correlate the noise to the original values because we want to gain as much data as possible from our limited resources. For example, for every value in the column "age", we increase or decrease it by one or two with a probability of 25\% each. This way the expectation value remains the same.

\begin{table}[h!tbp]
    \caption{Data set with added noise}
    \begin{center}
    \begin{tabular}{|c|c|c|c|c|}
    \hline
    \multicolumn{5}{|c|}{\textbf{Extract of Medical Record (Fictional)}} \\
    \cline{1-5}
    \textbf{\textit{Name}}& \textbf{\textit{Age}}& \textbf{\textit{Gender}}& \textbf{\textit{ZIP Code}} &\textbf{\textit{Diagnosis}} \\
    \hline
      Jane Doe& \sout{44} \textcolor{red}{43} & Female & 12345 & Cancer  \\
    \hline
        John Smith &\sout{22} \textcolor{red}{23} & Male & 12333 & Migraine \\
    \hline
        William Wonker &\sout{39} \textcolor{red}{37} & Male & 12344 & Incontinence\\
    \hline
     Harrison Seat &\sout{35} \textcolor{red}{37} & Male & 12355 & Incontinence\\
    \hline
     Bettina Wonker &\sout{42} \textcolor{red}{41} & Female & 12344 & No illness\\
    \hline
     Thomas Müller &\sout{22} \textcolor{red}{23} & Male & 12222 & Diabetes\\
    \hline
     Sharon Carter &\sout{47} \textcolor{red}{45} & Female & 12544 & Cancer\\
    \hline
     Maria Granger &\sout{27} \textcolor{red}{29} & Female & 12345 & Cancer\\
    \hline
     Christian Cloud & \sout{26} \textcolor{red}{25} & Male & 12333 & No illness\\
    \hline
     Kim Schmidt &\sout{21} \textcolor{red}{22} & Female & 12222 & Diabetes\\
    \hline
    \end{tabular}
    \label{tab4}
    \end{center}
\end{table}

If it is known to a potential attacker how the data has been altered, problems are caused similar to the ones discussed in the "Generalization and suppression" chapter. This is because we are basically creating another range the value can be in, but we reduce the inaccuracy a little bit. Noise addition can be used with random values which do not correlate to the original data as well, but the information loss then would be too high when applying it to every value. For a limited amount of values this could be possible, but then it is very difficult to balance privacy and information loss (even more difficult than it is already).\\

\subsubsection{Data swapping}
Like the name already states, when using this method, several values are swapped (now focusing on sensitive data!). This way the original sensitive information itself can be maintained and the participants privacy is preserved because its sensitive information could be that of another person, giving the participant plausible deniability. This technique can also be combined with frameworks such as the already described k-anonymity \cite{ldivers}. This method can be very useful when the sole amount of sensitive information is required (e.g., the amount of people taking illegal drugs). If this is not the case, it is not likely to be a suitable method for medical data because our data sets are often of a very limited size. For data swapping to be useful for medical data, the range in which the data is swapped should be controllable. Medical data often is correlated to age. Controlling the range would give us the opportunity to still learn the general contribution among the different ages (see section, "Rank swapping" for detailed information). The most critical parameter to adjust here is the number of values that are swapped. A number which is too low would not guarantee a suitable privacy for the participant because he probably would not be believed when he said that this was not his actual value. However, a number which is too high would make the results almost unusable for the researchers.\\

\begin{table}[h!tbp]
    \caption{Data set with swapped data}
    \begin{center}
    \begin{tabular}{|c|c|c|c|c|}
    \hline
    \multicolumn{5}{|c|}{\textbf{Extract of Medical Record (Fictional)}} \\
    \cline{1-5}
    \textbf{\textit{Name}}& \textbf{\textit{Age}}& \textbf{\textit{Gender}}& \textbf{\textit{ZIP Code}} &\textbf{\textit{Diagnosis}} \\
    \hline
      Jane Doe& 44 & Female & 12345 & \sout{Cancer} Incontinence  \\
    \hline
        John Smith & 22 & Male & 12333 & \sout{Migraine} Diabetes \\
    \hline
        William Wonker & 39 & Male & 12344 & Incontinence\\
    \hline
     Harrison Seat & 35 & Male & 12355 & \sout{Incontinence} Cancer\\
    \hline
     Bettina Wonker & 42 & Female & 12344 & No illness\\
    \hline
     Thomas Müller & 22 & Male & 12222 & Diabetes\\
    \hline
     Sharon Carter & 47 & Female & 12544 & Cancer\\
    \hline
     Maria Granger & 27 & Female & 12345 & Cancer\\
    \hline
     Christian Cloud & 26 & Male & 12333 & No illness\\
    \hline
     Kim Schmidt & 21 & Female & 12222 & \sout{Diabetes} Migraine\\
    \hline
    \end{tabular}
    \label{tab5}
    \end{center}
\end{table}

\subsubsection{Rank swapping}
Rank swapping, as well as data swapping, aims to minimize the information loss while anonymizing the data.
To apply rank swapping, all records of $X$ have to be sorted in increasing order of the values $x_{ij}$ of the attribute $j$ \cite{rankswapping}, which is to be anonymized.

\begin{table}[h!tbp]
    \caption{Data set sorted by age}
    \begin{center}
    \begin{tabular}{|c|c|c|c|c|}
    \hline
    \multicolumn{5}{|c|}{\textbf{Extract of Medical Record (Fictional)}} \\
    \cline{1-5}
    \textbf{\textit{Name}}& \textbf{\textit{Age}}& \textbf{\textit{Gender}}& \textbf{\textit{ZIP Code}} &\textbf{\textit{Diagnosis}} \\
    \hline
     Kim Schmidt & 21 & Female & 12222 & Diabetes\\
    \hline
    John Smith & 22 & Male & 12333 & Migraine \\
    \hline
     Thomas Müller & 22 & Male & 12222 & Diabetes\\
     \hline
     Christian Cloud & 26 & Male & 12333 & No illness\\
     \hline
     Maria Granger & 27 & Female & 12345 & Cancer\\
    \hline
    Harrison Seat & 35 & Male & 12355 & Incontinence\\
    \hline
     William Wonker & 39 & Male & 12344 & Incontinence\\
    \hline
     Bettina Wonker & 42 & Female & 12344 & No illness\\
    \hline
      Jane Doe& 44 & Female & 12345 & Cancer  \\
    \hline
     Sharon Carter & 47 & Female & 12544 & Cancer\\
    \hline
    \end{tabular}
    \label{tab6}
    \end{center}
\end{table}

After $X$ is sorted $x_{ij} \leq x_{lj}$ for all $1 \leq i < l \leq n$ applies for all the values in x (per definition). Every value $x_{ij}$ is swapped with $x_{lj}$. To adjust the area from which the value to be swapped can be (randomly) chosen, one can variate the parameter $p$, which limits the maximum distance $l$ can be apart from $i$ \cite{rankswapping}. This is because $l$ is defined to be in the range $i < l \leq i+p$. Simply speaking this means that each value is swapped with a value at a higher rank than itself, but not farther away than $p$ ranks.\\

\begin{table}[h!tbp]
    \caption{Data set with rank swapping of range 2 applied}
    \begin{center}
    \begin{tabular}{|c|c|c|c|c|}
    \hline
    \multicolumn{5}{|c|}{\textbf{Extract of Medical Record (Fictional)}} \\
    \cline{1-5}
    \textbf{\textit{Name}}& \textbf{\textit{Age}}& \textbf{\textit{Gender}}& \textbf{\textit{ZIP Code}} &\textbf{\textit{Diagnosis}} \\
    \hline
     Kim Schmidt & \sout{21} 22 & Female & 12222 & Diabetes\\
    \hline
    John Smith & \sout{22} 21& Male & 12333 & Migraine \\
    \hline
     Thomas Müller & \sout{22} 27 & Male & 12222 & Diabetes\\
     \hline
     Christian Cloud & \sout{26} 35& Male & 12333 & No illness\\
     \hline
     Maria Granger & \sout{27} 22 & Female & 12345 & Cancer\\
    \hline
    Harrison Seat & \sout{35} 26 & Male & 12355 & Incontinence\\
    \hline
     William Wonker & \sout{39} 42 & Male & 12344 & Incontinence\\
    \hline
     Bettina Wonker & \sout{42} 39 & Female & 12344 & No illness\\
    \hline
      Jane Doe& \sout{44} 47 & Female & 12345 & Cancer  \\
    \hline
     Sharon Carter & \sout{47} 44 & Female & 12544 & Cancer\\
    \hline
    \end{tabular}
    \label{tab7}
    \end{center}
\end{table}

Now that the order of the sensitive values is a completely different one, one could assume that the gathered information is useless; but due to the fact that we can adjust the range, in which the values are swapped, we can still determine the tendency of the result. For example, a simplified, fictional study gathers age, the amount of sexual partners a person had in his or her life and whether the person suffers (or had suffered) at least one sexual disease from 1000 participants. These records are sorted by the amount of sexual partners, and then swapped within a range limited by p = 7. Even though not a single value remains at the same place, the researchers come to the conclusion that the risk of getting a sexual disease increases with the amount of sexual partners a person had. The exact amount in each case was completely irrelevant for observing this tendency from the beginning; so the loss of information is neglectable in this example. On the downside, the fact that the range in which values can be swapped is fixed, and other information is not anonymized (at least not when using only this method), results in a vulnerability against reidentification attacks \cite{rankswapping}, where hackers use background information (or information gathered from other records) to reidentify a person in a database. Another problem which could occur is that people of the same age might have a higher probability of having the same disease, similar to what has been discussed at the "Generalization an suppression" section.\\

\subsubsection{Microaggregation}
Similar to the first method discussed in this chapter, microaggregation in general wants to maintain privacy; at least to the point that one individual can not be identified among a group of k participants (k-anonymity).\\

As before, all data set entries are sorted by the attribute we want to "anonymize" and divided into groups of size k. 

\begin{table}[h!tbp]
    \caption{Data set sorted into equvalence classes}
    \begin{center}
    \begin{tabular}{|c|c|c|c|c|}
    \hline
    \multicolumn{5}{|c|}{\textbf{Extract of Medical Record (fictional)}} \\
    \cline{1-5}
    \textbf{\textit{Name}}& \textbf{\textit{Age}}& \textbf{\textit{Gender}}& \textbf{\textit{ZIP Code}} &\textbf{\textit{Diagnosis}} \\
    \hline
    \multicolumn{5}{|c|}{Equivalence class one}\\
    \hline
      Jane Doe & 44 & Female & 12345 & Cancer  \\
    \hline
     Sharon Carter & 47 & Female & 12544 & Cancer\\
    \hline
     Bettina Wonker & 42 & Female & 12344 & No illness\\
    \hline
    \multicolumn{5}{|c|}{Equivalence class two}\\
    \hline
     Maria Granger & 27 & Female & 12345 & Cancer\\
    \hline
    Kim Schmidt & 21 & Female & 12222 & Diabetes\\
    \hline
    \multicolumn{5}{|c|}{Equivalence class three}\\
    \hline
        John Smith & 22 & Male & 12333 & Migraine \\
    \hline
    Christian Cloud & 26 & Male & 12333 & No illness\\
    \hline
     Thomas Müller & 22 & Male & 12222 & Diabetes\\
    \hline
    \multicolumn{5}{|c|}{Equivalence class four}\\
    \hline
        William Wonker & 39 & Male & 12344 & Incontinence\\
    \hline
     Harrison Seat & 35 & Male & 12355 & Incontinence\\
    \hline
    
    \end{tabular}
    \label{tab8}
    \end{center}
\end{table}

Now instead of generalizing them as before, each value is altered to the average value of the group \cite{microaggregation}. This microaggregation for one attribute (age for our interest most of the time) is called univariate aggregation \cite{microaggregation}. It has a big advantage over the generalization method (at least for purposes of analyzing medical data) because it decreases the divergence of our values, and therefore supplies more accuracy. (It still might be a problem depending on the specific study, but maybe a minor one now.) It also appears to be more suitable than the noise addition, because the range we draw the average from is unknown.\\

Instead of focusing on solely one attribute, multiple attributes can be addressed simultaneously which is then called multivariate aggregation. It proves to be difficult to form groups as in the univariate aggregation section because we have to consider multiple attributes now.
One way could be to concatenate the values to be considered. \cite{microaggregation}

Another way of accomplishing it would be by the use of appropriate distance vectors. (The specific functionality of these is not necessary to understand the following explanations). First, we determine the average record, as well as the most distant records to each side. As a result of using appropriate distance metrics, records close to each other should be very similar. Second, we form groups of size k around the most distant values. If there are more values left than can fit into two groups (more than 2k values), we continue to form groups as before. Otherwise, the remaining values form the last group \cite{microaggregation}.\\

In the table below (\ref{tab4}), the entries are sorted by gender and age, to achieve groups as similar as possible. Afterwards, the age is replaced by the average value.\\

\begin{table}[h!tbp]
    \caption{Data set with multivariate aggregation}
    \begin{center}
    \begin{tabular}{|c|c|c|c|c|}
    \hline
    \multicolumn{5}{|c|}{\textbf{Extract of Medical Record (fictional)}} \\
    \cline{1-5}
    \textbf{\textit{Name}}& \textbf{\textit{Age}}& \textbf{\textit{Gender}}& \textbf{\textit{ZIP Code}} &\textbf{\textit{Diagnosis}} \\
    \hline
    \multicolumn{5}{|c|}{}\\
    \hline
      Jane Doe & \sout{44} 44& Female & 12345 & Cancer  \\
    \hline
     Sharon Carter & \sout{47} 44 & Female & 12544 & Cancer\\
    \hline
     Bettina Wonker &\sout{42} 44 & Female & 12344 & No illness\\
    \hline
    \multicolumn{5}{|c|}{}\\
    \hline
     Maria Granger & \sout{27} 24 & Female & 12345 & Cancer\\
    \hline
    Kim Schmidt & \sout{21} 24 & Female & 12222 & Diabetes\\
    \hline
    \multicolumn{5}{|c|}{}\\
    \hline
        John Smith & \sout{22} 23 & Male & 12333 & Migraine \\
    \hline
    Christian Cloud & \sout{26} 23 & Male & 12333 & No illness\\
    \hline
     Thomas Müller & \sout{22} 23 & Male & 12222 & Diabetes\\
    \hline
    \multicolumn{5}{|c|}{}\\
    \hline
        William Wonker & \sout{39} 37 & Male & 12344 & Incontinence\\
    \hline
     Harrison Seat & \sout{35} 37 & Male & 12355 & Incontinence\\
    \hline
    
    \end{tabular}
    \label{tab9}
    \end{center}
\end{table}

Taking multiple attributes into consideration is important and especially interesting when thinking of side effects. For example, the vaccine AstraZeneca was not recommended for females under the age of 60 \cite{astra}; for male persons under this age, on the other hand, no significant risk could be detected.

\vspace*{0.6\baselineskip}
\subsection{Randomized response based privacy preserving data mining (PPDM)} Randomization alters the given data with a certain probability, which can be modified to match the occasion and needed accuracy. Not knowing whether the sent information is "real" data or modified, the privacy of each client can be guaranteed since even if the response is revealed, the client could still claim that the response must have been randomized \cite{google}. The differential privacy is depending on the probability of altering the results. Still, it has to be considered, that if the probability is higher, the result gets less accurate at the same time. Also a large group of participants is needed in order to get usable results. The RAPPOR algorithm discussed later on is based on randomization.

\vspace*{0.6\baselineskip}
\subsection{Cryptography based PPDM}
Since most of the following approaches are based on "Secure Multiparty Computation" (SMC), the fundamentals of this convention are explained in the following. SMC is a method to jointly create a function for their inputs while keeping their own input a secret. For instance, Alice, Bob and Charlie want to find out whether the majority of them are vaccinated or not. So, they decide to do a secret voting. Voting "1" means "I am vaccinated" and voting "0" means "I am not vaccinated". What they are trying to create is a function f(x) where f(0) equals the total amount of people being vaccinated in their group . To accomplish that, each of them creates a function (Alice: a(x), Bob: b(x), Charlie: c(x)) which results in their vote for x = 0 \cite{smc}. It is known that every function of degree k can be exactly defined by k+1 known points of this function. Since we have three participants, lets assume we are looking for a quadratic function. Alice therefore calculates three points: 
\begin{itemize}
    \item a(1), which she keeps to herself
    \item a(2), which she shares with Bob
    \item a(3), which she shares with Charlie
\end{itemize}
Each of the others behave similar, leading to Alice knowing a(1), b(1) and c(1), Bob knowing a(2), b(2) and c(2) and Charlie knowing a(3), b(3) and c(3). Since we defined $f(x) = a(x) + b(x) + c(x)$, each person knows f(x) for a certain x. Alice knows f(1) = a(1) + b(1) + c(1). So Alice shares f(1), Bob shares f(2) and Charlie shares f(3). With this information they can recreate f(x) using the Lagrange Interpolation \cite{smc}:\\
\begin{itemize}
    \item if h(x), a polynomial of degree at most $l$, and
    \item $|C| = l+1$, then it holds \\
    \begin{equation}
    h(X) = \sum_{i\in C} h(i)\delta_i(X)
\end{equation}
where 
\begin{equation}
    \delta_i(X) = \prod_{j\in C, j\neq i} \frac{x - j}{i - j}
\end{equation}
\end{itemize}
C in our case is $\{1,2,3\}$. Since Alice, Bob and Charlie chose their function, so that their function results in their vote for $x=0$ and $f(0) = a(0) + b(0) + c(0)$, f(0) now gives us the total number of vaccinated people in their group. During this process, each of them only knew one point of each others chosen function, which is not sufficient enough to find out their function, and thus how they voted \cite{smc}. Naturally, this requires the participants to behave truthfully and not collaborate with other members in terms to find out how they voted. Assuming Alice and Bob are vaccinated, Charlie is not and they all voted accordingly, the result f(0) = 2 tells us that there are two vaccinated people in this group. Charlie being the one who is not vaccinated now knows that the other two are vaccinated, which is unavoidable in this case. For Alice and Bob however, it is not possible to distinguish how they voted, which preserves their privacy.\\

The following strategies are (nearly) founded on SMC. Due to the amount of strategies, only short overviews are given together with the source to read for detailed information of the mechanics.\\

\subsubsection{Kantarcioglu and Clifton} Kantarcioglu and Clifton \cite{KantarciogluClifton} developed a strategy for affiliation rule mining by fusing different cryptographic strategies. The information has to be evenly apportioned. The strategy sets a limitation to shared data while avoiding large overheads to the mining task. \cite{currentMethods}\\

Affiliation rules, in general, describe the co-occurrence of events beyond randomness. Simply speaking, an association rule can be compared to an if-then-condition: $A \rightarrow B$. A and B are both a set of things (or events) that are contained in $I$, but do not share any elements ($A \subset I, B \subset I$ and $A \cap B = \varnothing$).\\

Without limitations there could be an tremendous number of co-occurrences, and therefore association rules. Consequently the amount of useful or \textit{solid affiliation rules} are limited by the condition that they have a \textit{support} limit and a base certainty edge. The support describes the frequent occurrence of A or B and can be used to filter out events or items that are not of great importance due to their rareness. A customary threshold for the support is 35\% \cite{affiliation}. The certainty describes how often A and B occur together and is a parameter to determine the randomness of the co-occurrence since co-occurrences are not necessarily correlations as well. 
\begin{equation}
    \begin{split}
    certainty(A \rightarrow B) &= P(B|A) 
    = \frac{support(A \cup B}{support(A)}\\
    &= \frac{support -  count(A \cup B)}{support - count(A)}
    \end{split}
\end{equation}
A customary threshold for certainty is 60\% \cite{affiliation}.
Affiliation rule mining usually deals with finding all regular item sets and creating solid affiliation rules.\\

Their approach is not dependent on a trusted third party server. Also, because of the simple random number addition used to calculate the support count of item sets, only a low computation cost is incurred \cite{improvingHealthcare}. On the downside, the information sent or received by targeted participants can be traced; and thus, the private value can be revealed by computing the difference of these values, violating privacy requirements \cite{improvingHealthcare}.\\
    
\subsubsection{Yang et al.} Yang et al. ensured a private access to their own information (and only their own information) with their strategy over evenly divided information. \cite{currentMethods}\\
    
\subsubsection{Vaidya and Clifton} Vaidya and Clifton mainly focused on the maximum security of affiliation rule digging and also suggested a specific strategy for grouping over vertically apportioned data\cite{currentMethods} \cite{vadiyaClifton}.

\vspace*{0.6\baselineskip}
\subsection{Further overview of existing approaches}

\subsubsection{Yi Huangs} Yi Huangs approach \cite{huang} is of interest because of its specific relation to healthcare data. It also deals with affiliation rule mining, but specifically on medical examination data and outpatient medical records. Using this approach, correlations between disease and abnormal test results can be detected \cite{improvingHealthcare}. In this case the data of a regional Taiwan Hospital was used, but privacy issues during the integration of the data were not taken into consideration.\\

\subsubsection{M. Hussein et al.}
Instead of a random number scheme, M. Hussein et al. \cite{hussein} used Paillier homomorphic encryption. This preserves the privacy against external as well as internal attackers, but only if the communication environment is secure. Otherwise the initiator finds the support count of the targeted site by tracing and disrupting the communication between combiner and normal participants \cite{improvingHealthcare}.\\

\subsubsection{Nanavati et al.}
Nanavati deals with privacy preserving affiliation rule mining on horizontally partitioned data using Shamir's secret sharing technique \cite{nanavati} (which is also the foundation of the already explained Secure Multiparty Computation). The count of all (candidate) item sets is computed, thus causing a high computation cost for the algorithm. As well as Hussein's approach, this method requires a secure communication environment among the collaborative participants.

\vspace*{0.6\baselineskip}
\section{RAPPOR}
RAPPOR is an acronym for "Randomized Aggregatable Privacy Preserving Ordinal Response". As stated in its name, it attempts to ensure the users (or in this case rather the patients) privacy as explained in the differential privacy section above.\\

RAPPOR is an open source project initiated and used by google chrome to collect user data \cite{chrome}. Google can use this data, e.g., to identify the cause of computer crashes or to review new design features, all while keeping the individuals users data a secret. In opposite to most other methods, the data is already anonymized before it is send to google, which is another indicator for strong privacy measures \cite{chrome}. Since RAPPOR is an open source project, you can review the mechanisms by yourself without needing to trust another party.\\

To explain its usage and mechanisms, the following example will be used:\\
Since sexual transmitted diseases (STDs) continue to be constantly tabooed in society, a researcher attempts to better estimate the underreported cases, of such, by questioning a large amount of people using the RAPPOR algorithm to ensure their privacy. She also wants to know which exact STD it is; in addition she gathers information on other illnesses in order to make possible correlations with other diseases (such as cancer) visible.

\subsection{The algorithm in detail}
Here are the given data and parameters to begin with:
\begin{itemize}
    \item \textbf{The client's value \textit{v}} - contains the unfiltered information that is to be sent to the aggregator. In this example this might be "chlamydia", "syphilis" or "none", stating the STD of the patient.
    \item A \textbf{Bloom filter \textit{B}} of size \textbf{\textit{k}} - to be further explained below.
    \item \textbf{\textit{h} hash functions} - multiple hash functions which aim to achieve diverging outcomes at the same input.
\end{itemize}
The first step of the algorithm is to hash \textit{v} onto \textit{B} using the hash functions. The Bloom filter can be illustrated as an array containing only zeros.
An example:
\begin{table}[h!tbp]
    \caption{Empty Bloom Filter B of size 12}
    \begin{center}
    \begin{tabular}{|c|c|c|c|c|c|c|c|c|c|c|c|}
    11&10&9&8&7&6&5&4&3&2&1&0 \\
    \hline
    0&0&0&0&0&0&0&0&0&0&0&0\\
    \hline
    \end{tabular}
    \label{bloomEmpty1}
    \end{center}
\end{table}

Assuming two given hash functions h1 and h2 and two given values "chlamydia" and "syphilis", we now depict a use case. In this use case, by inserting "chlamydia", h1 equates to three and h2 equates to eleven, resulting to the following array: 

\begin{table}[h!tbp]
    \caption{Bloom Filter B after first value}
    \begin{center}
    \begin{tabular}{|c|c|c|c|c|c|c|c|c|c|c|c|}
    11&10&9&8&7&6&5&4&3&2&1&0 \\
    \hline
    \textcolor{red}{1}&0&0&0&0&0&0&0&\textcolor{red}{1}&0&0&0\\
    \hline
    \end{tabular}
    \label{bloomEmpty2}
    \end{center}
\end{table}

The insertion of "syphilis" leads to the values four for h1 and eight for h2 leading to the following:

\begin{table}[h!tbp]
    \caption{Bloom Filter B after second value}
    \begin{center}
    \begin{tabular}{|c|c|c|c|c|c|c|c|c|c|c|c|}
    11&10&9&8&7&6&5&4&3&2&1&0 \\
    \hline
    \textcolor{red}{1}&0&0&\textcolor{orange}{1}&0&0&0&\textcolor{orange}{1}&\textcolor{red}{1}&0&0&0\\
    \hline
    \end{tabular}
    \label{bloomEmpty3}
    \end{center}
\end{table}

If the bloom filter is now is tested for a yet not hashed value "aids", which results in the values eleven and four, we would assume that this value has been hashed onto the bloom filter before. However, this is not true int this case. Cases in which we assume something even though it is not true are called "false positive". As can be seen from this false positive, even if an attacker would somehow know B, he or she still could not be completely sure to make assumptions about the client because the multiple hash functions cause uncertainty to the origin of the set bits.\\

The second step is the \textbf{permanent randomized response} (PRR). Additional parameters are: 
\begin{itemize}
    \item \textbf{\textit{f}} - a "user-tunable parameter controlling the level of longitudinal privacy guarantee" \cite{google}.
\end{itemize}
The algorithm of the permanent randomized response is defined as follows: "For each client's value \textit{v} and bit \textit{i}, $0 \leq i < \textit{k}$ in \textit{B}, create a binary reporting value $B'_i$ which equals to \\
\begin{equation}
B'_i =
\begin{cases} 
                1, &\text{with probability} \frac{1}{2}f  \nonumber\\
                 0, &\text{with probability} \frac{1}{2} f \nonumber\\
                 B_i &\text{with probability} $1-f$ \nonumber
\end{cases} \\
\end{equation}
Subsequently, this \textit{B'} is memorized and reused as the basis for all future reports on this distinct value \textit{v}" \cite{google}.\\

This randomization method can be interpreted as two consecutive coin-flips. An example: The patient is asked whether he or she had a sexual transmitted disease in the past, in order to obtain data on a statistical significant coherence of another disease emerging. Before sending the answers the user has given, a coin is flipped. On tails, the given answer is transmitted. Otherwise, a random answer is transmitted (determined by another coin flip). This way when having a large amount of people participating in the study and knowing the chances of the answer being altered, the amount of people with sexual transmitted diseases can still be determined with good accuracy. Still the individual has a reasonable deniability because  he or she can still claim that the given answer has been altered.\\

Instead of altering the answer directly, the bits contained in the bloom filter are altered (or not). Notice that the first "coin-flip" does not have to be equally balanced. By tuning \textit{f}, one can influence the probability of how likely it is that the bit has changed. However, even if the bit gets changed, it could lead to the same outcome as if it hasn't been changed. Therefore the probability of the bit staying the same is $1-\frac{1}{2}f$. The higher f is, the higher the chance of the client's value will be altered; therefore, a large differential privacy can be ensured. At the same time, a high f also makes it more difficult for the researchers to interpret the results correctly. Tuning this parameter to the needs and possibilities is therefore of utmost importance. Also \textit{B'} has to be used for all future reportings on the information about \textit{B}. Otherwise a potential attacker would be able to estimate \textit{B} by observing multiple noisy versions of it\cite{google}.\\

Before sending the generated report to the server, the \textbf{instantaneous randomized response} (IRR) has to be applied. The following additional parameters are needed:
\begin{itemize}
    \item \textbf{\textit{S}} - the generated report. It is of size \textit{k} (the same size as the bloom filter).
    \item \textbf{\textit{q}} - $0 < q < 1$, parameter to tune the likeliness of a bit i getting set to one, given that the bit was one before.
    \item \textbf{\textit{p}} - $0 < p < 1$, parameter to tune the likeliness of a bit i getting set to one, given that the bit wasn't one before.
\end{itemize}S is initialized to zero. After the initialization each bit \textit{i} in \textit{S} is set with the following probabilities: 

\begin{equation}
P(S_i = 1) =
\begin{cases}
     & q, if B'_i = 1. \nonumber\\
     & p, if B'_i = 0.

\end{cases}
\end{equation}
The IRR makes it difficult to track a client (otherwise \textit{B'} could be used to identify multiple reportings of the same participant).

After these applications, the report \textit{S} can be sent to the server.

\subsection{Differential privacy of RAPPOR}
The differential privacy $\epsilon$ of the PRR can be defined as follows (as proven in the paper of Google regarding RAPPOR \cite{google}):
\begin{equation}
    \epsilon_\infty = 2h \ln \Bigg(\frac{1-\frac{1}{2}f}{\frac{1}{2}f}\Bigg)
\end{equation}

The equation for the differential privacy regarding the IRR requires a lemma:\\
LEMMA 1. Probability of observing "1" given that the underlying bloom filter bit was set is given by
\begin{equation}
    q* = P(S_i = 1| b_i = 1) = \frac{1}{2}f(p+q) + (1-f)q 
\end{equation}
Probability of observing "1" given that the underlying Bloom filter bit was not set is given by
\begin{equation}
    p* = P(S_i = 1| b_i = 0) = \frac{1}{2}f(p+q) + (1-f)p  
\end{equation}
This being defined, the differential privacy $\epsilon_1$ can be calculated as follows:
\begin{equation}
    \epsilon_1 = h \log \Bigg(\frac{q*(1-p*)}{p*(1-q*)}\Bigg)
\end{equation}\\
 A calculation for illustration purposes:
 \begin{itemize}
     \item h: 2
     \item f: 50\% (0.5)
     \item q: 75\% (0.75)
     \item p: 50\% (0.5)
 \end{itemize}
 The choice of these parameters would lead to the following values:\\
 permanent randomized response
 \begin{equation}
 \begin{split}
    \epsilon_\infty = 2h \ln \Bigg( \frac{1-\frac{1}{1}f}{\frac{1}{2}f}\Bigg)\\
    = 4 \ln \Bigg( \frac{0.75}{0.25} \Bigg)\\
    \thickapprox 4.3945
\end{split}
\end{equation}
 instantaneous randomized response
\begin{equation}
\begin{split}
    \epsilon_1 = h \log \Bigg( \frac{q*(1-p*)}{p*(1-q*)}\Bigg)\\
    \thickapprox 1.5499
\end{split}
\end{equation}

\vspace*{0.6\baselineskip}
\subsection{RAPPOR in CrypTool 2}
Before coming to a conclusion, I would like to introduce a RAPPOR template for learning purposes, which can be found in CrypTool 2 (download: https://www.cryptool.org/de/ct2/downloads). It is structured into five tabs:
\begin{itemize}
    \item "Start"
    \item "Overview"
    \item "Bloom Filter"
    \item "Randomized Response"
    \item "Heat Map"
\end{itemize}

Using this tool, you can adjust all the described parameters (number of hash functions, bloom filter size, etc.). The "Start" tab gives a short tutorial on how to navigate through the template and on how to adjust the different parameters. In the "Overview" tab it shows the altering of the boolean array in the process of applying the permanent randomized response as well as the instantaneous randomized response several times (according to the settings). The "Bloom Filter" tab gives a detailed insight on how the empty bloom filter is filled with values, according to the data set of the user. The heat map visualizes how often a cell of the array has been set (explanations of the different tabs and visualizations can also be found at the bottom of each tab). The most interesting tab (in my personal opinion) is the "Randomized Response" tab because it visualizes and explains all the possibilities of the outcomes (and therefore, answers the question \textit{How likely is it, that the original answer is submitted?}). It also shows the differential privacy levels and how they are calculated.

\begin{figure}[htbp]
\centerline{\includegraphics[scale = 0.42]{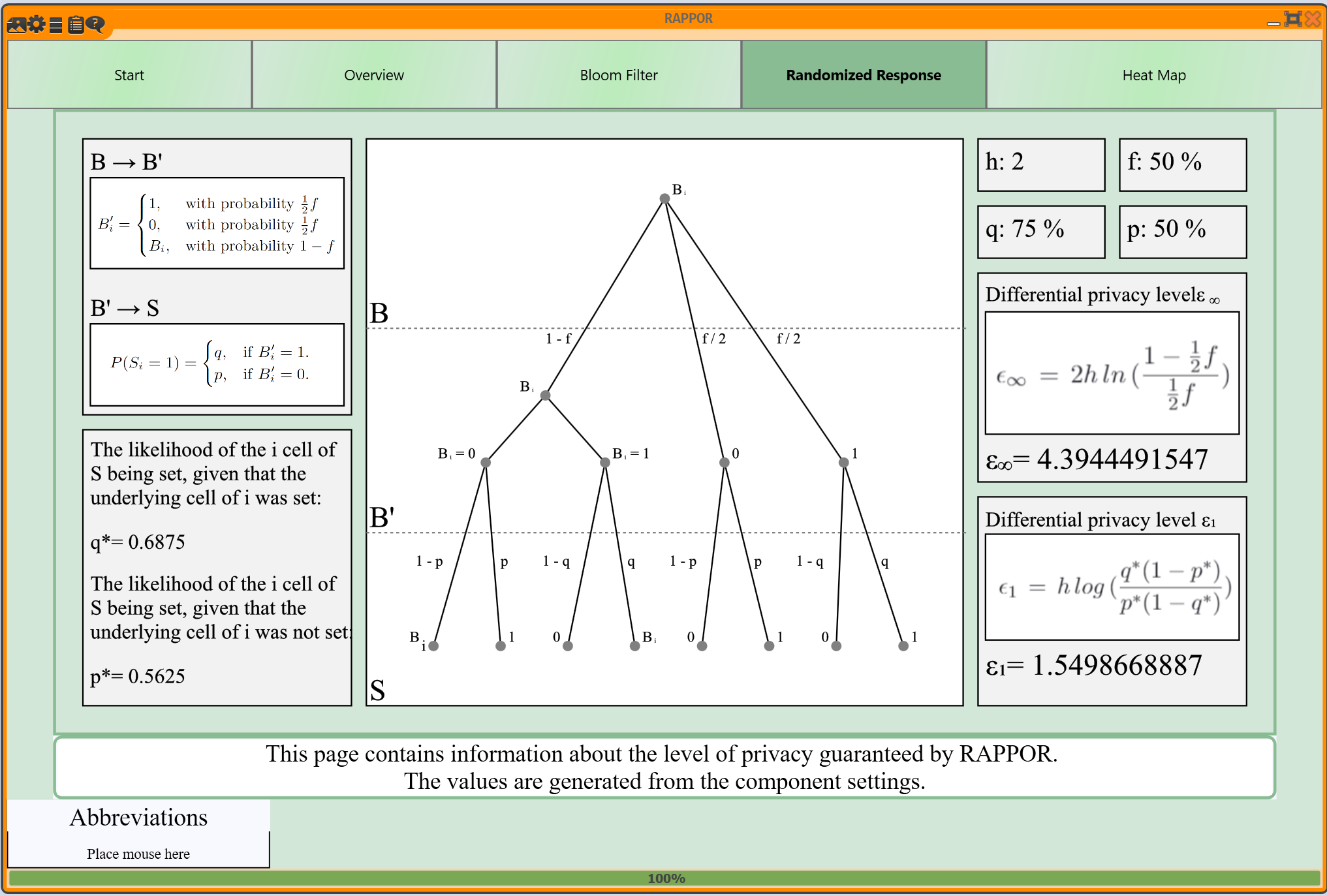}}
\caption{"Randomized Response" Tab in CrypTool 2.}
\label{fig}
\end{figure}

\vspace*{0.6\baselineskip}
\section{Conclusion}
Anonymization in context of medical data is particularly difficult, considering the high sensitivity of the information (as always in combination with its usefulness for a researcher). The "Generalization and suppression" method is the one with the most shortcomings in my opinion. Due to issues addressed by k-anonymity and l-diversity, the required group size remains high, but without optimizing the information loss. Although, for obvious reasons, it is a standard to participate in a field study "anonymously" (meaning that the name is suppressed); that does not nearly guarantee true anonymity without any other methods applied. Further manipulation of the data comes with a loss of information, which often makes the results useless. Both k-diversity and l-diversity address the privacy problem, but do not have an impact (or at least not a positive one) on the information loss. The information loss might be ignorable when collecting data of overall behaviour resulting in health issues, where the group size is much larger (e.g., smoking). In this case, generalization is a very simple and easy implementable method. In general, I do not think generalization and suppression should be viewed as standalone methods, but as a basic function to build upon.\\

Noise addition and rank swapping struggle with similar problems, but to a smaller degree. They seem more suitable in order to balance privacy and information loss, because the parameters (range in this case) can be easily adjusted and the range is not as obviously detectable. With microaggregation, one does not know the range at all which is a huge improvement to the other methods. However, it is not as easily implemented because it can be difficult to find suitable groups to draw the average from.\\

The RAPPOR algorithm is a strong method suitable for the aggregation of medical data, if a large enough group of people participate. It can be a suitable measure against several kinds of attacks on the data due to the tunable variables. One of its benefits is that original data can be used (instead of the altered data, which loses information during anonymization), which is often a requirement for the researches of some critical diseases. Each method has its strength and weaknesses and is important to contribute to privacy preserving measures, for not every problem can be solved with the same solution to the same contentment. RAPPOR especially persuades due to its versatility because of the many adjustable parameters. But as always, the group size is critical here in order to choose the right method. Many of these methods do not suffice alone, but can accomplish a solid result when being combined with another.

\vspace*{0.6\baselineskip}


\begin{thebibliography}{00}
\bibitem{currentMethods} S. Joseph Gabriel and Dr. P. Sengottuvelan, "A Survey On Privacy Preserving Data Mining Its Related Applications In Health Care Domain", European Journal of Molecular \& Clinical Medicine, vol. 07, Issue 08, 2020

\bibitem{governor} arsTECHNICA. \enquote{"Anonymized" data really isn't - and here's why not}.  https://arstechnica.com/tech-policy/2009/09/your-secrets-live-online-in-databases-of-ruin/ accessed on 30.05.2022

\bibitem{google} Erlingsson, Úlfar, Vasyl Pihur, and Aleksandra Korolova.
"Rappor: Randomized aggregatable privacy-preserving ordinal response".
Proceedings of the 2014 ACM SIGSAC conference on computer and communications security. 2014.
https://dl.acm.org/doi/abs/10.1145/2660267.

\bibitem{dataRegulation} Datenschutz.org.
"Vergleich von Datenschutzrichtlinie und EU-Grundverordnung zum Datenschutz".
\url{https://www.datenschutz.org/eu-datenschutzgrundverordnung/}

\bibitem{privacy}Cynthia Dwork: Differential Privacy. In: 33rd International Colloquium on Automata, Languages and Programming, part II (ICALP 2006). Springer, Juli 2006, S. 1–12, "doi:10.1007/11787006\_1".

\bibitem{affiliation} Odule, Tola J. Adesina, Ademola O. Abdullah, K-K. Adebisi, Ogunyinka, Peter I. "Using Affiliation Rules-based Data Mining Technique in Referral System" Iraqi Journal of Science, 2020, Vol. 61, No. 11, pp: 3095-3103 DOI: 10.24996/ijs.2020.61.11.30

\bibitem{KantarciogluClifton}
Kantarcioglu,M., \& Clifton, C. (2004). Privacy-preserving distributed mining of association rules on horizontally partitioned data. IEEE transactions on knowledge and data engineering, 16(9), 1026-1037.

\bibitem{vadiyaClifton}Vaidya, J., \& Clifton, C. (2003, August). Privacy-preserving k-means clustering over vertically partitioned data. In Proceedings of the ninth ACM SIGKDD international conference on Knowledge discovery and data mining (pp. 206-215).

\bibitem{improvingHealthcare} Nikunj Domadiya, Udai Pratap Rao, (2020). Improving healthcare services using source anonymous
scheme with privacy preserving distributed healthcare data collection and mining. https://doi.org/10.1007/s00607-020-00847-0

\bibitem{huang}Huang Yi(2013) Mining association rules between abnormal health examination results and outpatient
medical records. Health Inf Manag J 42(2):23

\bibitem{hussein}Hussein M, El-Sisi A, Ismail N (2008) Fast Cryptographic Privacy Preserving Association Rules
Mining on Distributed Homogeneous Data Base. Knowledge-Based Intelligent Information and Engineering
Systems vol 5178, pp 607–616. https://doi.org/10.1007/978-3-540-85565-1\_75


\bibitem{nanavati}Nanavati NR, Lalwani P, Jinwala DC (2014) Analysis and evaluation of schemes for secure sum incollaborative frequent item set mining across horizontally partitioned data. J Eng 2014:110

\bibitem{ldivers}Charu C. Aggarwal, Philip S. Yu (2008) A GENERAL SURVEY OF PRIVACY-PRESERVING DATA MINING MODELS AND ALGORITHMS, chapter 2

\bibitem{rankswapping} Javier Herranz, Jordi Nin (2014) Secure and efficient anonymization of distributed confidential
databases, chapter 2.3

\bibitem{kAnonymity}Latanya Sweeney: k-anonymity: A model for protecting privacy In: International Journal of Uncertainty, Fuzziness and Knowledge-Based Systems, Vol. 10, Issue 5, World Scientific, 2002, S. 557–570 (english).

\bibitem{astra}Robert Koch Institut, "Pressemitteilung der STIKO zum AstraZeneca-Impfstoff", https://www.rki.de/DE/Content/Kommissionen/STIKO/Empfehlungen/A
straZeneca-Impfstoff-2021-03-30.html

\bibitem{bloom}Rob Edwards, Bloom Filters https://youtu.be/heEDL9usFgs

\bibitem{microaggregation}Josep Domingo-Ferrer, Jordi Soria-Comas, David Sánchez, Database Anonymization: Privacy Models, Data Utility, and
Microaggregation-based Inter-model Connections, ISBN: 9781627058445

\bibitem{gendermedicine}muenchen-klinik, https://www.muenchen-klinik.de/gendermedizin-frau/

\bibitem{vorsorge}krankenkasseninfo.de, https://www.krankenkasseninfo.de/vorsorgeunters
uchungen/

\bibitem{diabetes}bundesgesundheitsministerium, https://www.bundesgesundheitsministeri
um.de/themen/praevention/gesundheitsgefahren/diabetes.html

\bibitem{gensur}Sibghat Ullah Bazai, Julian Jang-Jaccard, In-Memory Data Anonymization Using Scalable and High Performance RDD Design

\bibitem{smc}CryptoClear, Basics of Secure Multiparty Computation, https://youtu.be/\_mDlLKgiFDY

\bibitem{chrome}Thorsten Schröder, Google schützt Nutzer beim Datensammeln, https://www.golem.de/news/rappor-google-schuetzt-nutzer-beim-datensammeln-1411-110259.html
\end{thebibliography}
\end{document}